\documentclass[a4paper,12pt]{article}
\usepackage[utf8x]{inputenc}
\usepackage{graphicx}
\usepackage[T1]{fontenc}
\usepackage{amsmath}
\usepackage{amsfonts}
\usepackage{amssymb}
\usepackage{amsthm}
\usepackage{amstext}
\usepackage{pstricks}

\newdimen\CdotAxis
\newcommand*{\CdotAux}[3]{%
  {%
    \settoheight\CdotAxis{$#2\vcenter{}$}%
    \sbox0{%
      \raisebox\CdotAxis{%
        \scalebox{#1}{%
          \raisebox{-\CdotAxis}{%
            $\mathsurround=0pt #2#3$%
          }%
        }%
      }%
    }%
    \dp0=0pt %
    \sbox2{$#2\bullet$}%
    \ifdim\ht2<\ht0 %
      \ht0=\ht2 %
    \fi
    \sbox2{$\mathsurround=0pt #2#3$}%
    \hbox to \wd2{\hss\usebox{0}\hss}%
  }%
}

\newtheorem{theorem}{Theorem}[section]

\theoremstyle{definition}

\theoremstyle{remark}

\def \cP {\mathcal{P}}

\def \cS {\mathcal{S}}

\def \a {\alpha}

\def \d {\delta}
\def \e {\varepsilon}

\def \l {\lambda}

\def \N {\mathbb{N}}

\def \R {\mathbb{R}}

\def \lra {\longrightarrow}

\def \pml {\cP^m_{\ell +1}}
\def \pmn {\cP^m_{N}}

\def\uro{\smash{{U}^{\!\!\!\!\raise5pt\hbox{$\scriptstyle o$}}}}

\begin{document}

\begin{center}
\begin{LARGE}
Convergence of a Moran model\\
to Eigen's quasispecies model
\end{LARGE}

\begin{large}
Joseba Dalmau

\vspace{-12pt}
Universit\'e Paris Sud and ENS Paris

\vspace{4pt}
\today
\end{large}
\end{center}

\begin{abstract}
\noindent
We prove that a Moran model
converges in probability 
to Eigen's quasispecies model
in the infinite population limit.
\end{abstract}

\section{Introduction}
The concept of quasispecies was proposed by Manfred Eigen
in order to explain how a population of macromolecules behaves 
when subject to an evolutionary process with selection and mutation.
In his celebrated paper~\cite{Eigen1},
Eigen models the evolution of a population of macromolecules
via a system of differential equations,
which arises from the laws of chemical kinetics.
Selection is performed according to a fitness landscape,
and mutations occur in the course of reproductions,
independently at each locus with rate $q$.
On the sharp peak landscape
---all but one sequence, the master sequence,
have the same fitness and the master sequence has higher fitness than the rest---
Eigen discovered that an error threshold phenomenon takes place:
there exists a critical mutation rate $q^*$
such that if $q>q^*$ then at equilibrium the population is totally random,
while if $q<q^*$ then at equilibrium the population forms a quasispecies,
i.e., it contains a positive fraction of the master sequence along with a cloud of mutants
that closely resemble the master sequence.
The concepts of error threshold and quasispecies might not only be relevant in molecular genetics,
but also in several other areas of biology, namely population genetics or virology~\cite{Domingo}.
Nevertheless, in Eigen's model the dynamics of the concentrations of the different genotypes
is driven by a system of differential equations,
which is a major drawback for the viability of the model in settings more complex than the molecular level~\cite{Wilke}.
A finite and stochastic version of Eigen's quasispecies model would be much more suitable
to expand the quasispecies theory to other areas~\cite{EMS,Schuster1,Wilke}.

The issue of designing a finite population version of the quasispecies model
has been tackled by several authors.
Different approaches have been considered in the literature:
Alves and Fontanari~\cite{AF}
propose a finite population model 
and they study the dependence of the error threshold on the population size,
a similar approach is taken by McCaskill~\cite{McCaskill},
Park, Mu\~noz, Deem~\cite{PMD}
and Saakian, Deem, Hu~\cite{SDH},
who all suggest different kinds of finite population models.
In~\cite{NS}, Nowak and Schuster derive the error threshold for finite populations
using a birth and death chain. 
More recently, in~\cite{CerfM,CerfWF},
Cerf shows that the error threshold and quasispecies concepts
arise for both the Moran model and the classical Wright--Fisher model
in the appropriate asymptotic regimes.
Some other authors propose stochastic models that converge to Eigen's model
in the infinite population limit,
this is the approach taken by Demetrius, Schuster, Sigmund~\cite{DSS},
who use branching processes,
Dixit, Srivastava, Vishnoi~\cite{DSV}
or Musso~\cite{Musso}.

Showing convergence of a finite population model
to Eigen's model is in general a delicate matter;
to our knowledge, all the works that have been done in this direction
prove that some stochastic process converges to Eigen's model in expectation.
As pointed out in~\cite{DSS}, convergence in expectation can be misleading sometimes,
mainly due to the fact that variation might increase as expectation converges,
leading to a poor understanding of the asymptotic behaviour of the stochastic process.
In the current work we consider the Moran model studied in~\cite{CerfM,CD} 
driving the evolution of a finite population subject to selection and mutation effects.
This Moran model is shown to converge to Eigen's quasispecies model in the infinite population limit,
independently of the fitness landscape:
on any finite time interval, we prove convergence in probability
for the supremum norm.
The interest of our result not only lies on the type of convergence,
but also on the choice of the model:
the Moran model is possibly one of the simplest models
for which such a result can be expected.
The result is proven by means of a theorem due to Kurtz~\cite{Kurtz},
which gives sufficient conditions for the convergence of a sequence
of Markov processes to a deterministic trajectory,
characterised by a system of differential equations.

The article is organised as follows:
first we briefly introduce Eigen's quasispecies model
and the Moran model.
We state the main result in section~\ref{Main}.
In section~\ref{Convmark} we adapt Kurtz's theorem,
which originally deals with continuous state space Markov chains,
to the discrete state space setting.
Finally, in section~\ref{Proof},
we apply Kurtz's theorem in order to prove the main result.

\section{The Eigen and Moran models}\label{Models}
We present here Eigen's quasispecies model
and a discrete Moran model.
Consider a set of $N$ different genotypes,
labelled from 1 to $N$.
Both Eigen's model and the Moran model
describe the evolution of a population
of individuals having genotypes $1,\dots,N$.
In both models the evolution of the population
is driven by two main forces:
selection and mutation.
The selection and mutation mechanisms depend only on the genotypes,
and are common to both models.
Selection is performed with a fitness landscape
$(f_i)_{1\leq i\leq N}$,
$f_i$ being the reproduction rate of an individual having genotype $i$.
The mutation scheme is encoded in a mutation matrix
$(Q_{ij})_{1\leq i,j\leq N}$,
$Q_{ij}$ being the probability that an individual having genotype $i$
mutates into an individual having genotype $j$.
The mutation matrix is assumed to be stochastic,
i.e., its entries are non--negative and the rows add up to 1.

\textbf{Eigen's model.} 
Eigen originally formulated the quasispecies model
to explain the evolution of a population of macromolecules.
The evolution of the concentration of the different genotypes
is driven by a system of differential equations,
obtained from the theory of chemical kinetics.
Let us denote by $\cS_N$ the unit simplex, i.e.,
$$\cS_N\,=\,
\lbrace\,
x\in\R^N:
x_i\geq 0,\ \,  1\leq i\leq N\quad
\text{and}\quad
x_1+\cdots+x_N=1
\,\rbrace\,.$$
An element $x\in\cS_N$ represents a population 
in which the concentration of the individuals having the $i$--th genotype
is $x_i$, for $1\leq i\leq N$.
Let $x^0\in\cS_N$ be the starting population
and let us denote by $x(t)$ the population at time $t>0$.
Eigen's model describes the dynamics of $x(t)$ 
thorough the following system of differential equations:
$$
(*)\quad
x_i'(t)\,=\,\displaystyle\sum_{k=1}^N f_k Q_{ki} x_k(t)
-x_i(t)\sum_{k=1}^N f_k x_k(t)\,,\quad 1\leq i\leq N\,,
$$
with initial condition $x(0)=x^0$.
The first term in the differential equation
accounts for the replication rate and mutations towards the $i$--th genotype,
while the second term helps to keep the total concentration constant.
A recent review on Eigen's quasispecies model can be found in~\cite{Schuster2}.

\textbf{The Moran model.}
Moran models aim at describing the evolution of a finite population.
The dynamics of the population is stochastic,
the evolution is described by a Markov chain.
Loosely speaking,
the Moran model evolves as follows:
at each step of time,
an individual is selected from the current population 
according to its fitness,
this individual then produces an offspring,
which is subject to mutations.
Finally,
an individual chosen uniformly at random from the population is replaced by the offspring.
The state space of the Moran process
will be the set $\pmn$ of the ordered partitions of the integer $m$ 
in at most $N$ parts:
$$\pmn\,=\,\lbrace\,
z\in\N^N:
z_1+\cdots+z_N=m
\,\rbrace\,.$$
An element $z\in\pmn$
represents a population in which $z_i$ individuals 
have the genotype $i$, for $1\leq i\leq N$.
The only allowed changes at each time step
consist in replacing an individual from the current population by a new one.
If we denote by $(e_i)_{1\leq i\leq N}$
the canonical basis of $\R^N$,
the only allowed changes in a population are of the form
$$z\,\lra\,z-e_i+e_j\qquad 1\leq i,j\leq N\,.$$
Let $\l$ be a constant such that
$\l\,\geq\,\max\lbrace\,f_i:1\leq i\leq N\,\rbrace$.
The Moran process is the Markov chain
$(Z_n)_{n\geq 0}$
having state space $\pmn$
and transition matrix $p$ given  by:
for all $z\in\pml$ and $i,j\in\lbrace\,1,\dots,N\,\rbrace$ such that $i\neq j$,
$$p(z,z-e_i+e_j)\,=\,
\frac{z_i}{m}\times\frac{1}{\l m}\sum_{k=1}^N f_kQ_{kj}z_k\,.$$
The other non--diagonal coefficients of the transition matrix are null,
the diagonal coefficients are arranged so that the matrix is stochastic,
i.e., the entries are non--negative and the rows add up to 1.

\section{Main result}\label{Main}
Our aim is to show that Eigen's quasispecies model
arises as the infinite population limit of the Moran model.
More precisely, we will prove the following result:
\begin{theorem}\label{main}
Let $(Z_n)_{n\geq0}$ 
be the Moran process described above.
Suppose that we have the convergence of the initial conditions
towards $x^0$:
$$\lim_{m\to\infty}\frac{1}{m}Z_0=x^0\,,$$
and let
$x(t)$ be the solution of the system of differential equations $(*)$
with initial condition $x(0)=x^0$.
Then, for every $\d,T>0$, we have
$$\lim_{m\to\infty}\,
P\bigg(
\sup_{0\leq t\leq T}\bigg|
\frac{1}{m}Z_{\lfloor\l mt\rfloor}-x(t)
\bigg|>\d
\bigg)\,=\,0\,.$$
\end{theorem}
This result is an immediate consequence of theorem~4.7 in~\cite{Kurtz}.
In order to prove the result, we proceed in two steps.
We state first theorem~4.7 in~\cite{Kurtz},
and we show next that all the hypotheses needed to apply the theorem
are fulfilled in our particular setting.

\section{Convergence of a family of Markov chains}\label{Convmark}
Let $d\geq 1$ and let $E$ be a subset of $\R^d$.
Let 
$\big(
(X^m_n)_{n\geq 0},
m\geq 1
\big)$
be a sequence of discrete time Markov chains 
with state spaces $E_m\subset E$
and transition matrices $(p^m(x,y))_{x,y\in E_m}$.
Let $F:\R^d\to\R^d$ and consider
the system of differential equations
$$x_i'(t)\,=\,F_i(x(t))\,,\qquad 1\leq i\leq N\,.$$
Theorem~$4.7$ in~\cite{Kurtz}
gives a series of sufficient conditions
under which the sequence of Markov chains $(X^m)_{m\geq 1}$
converges to a solution of the above system of differential equations.
The original statement of theorem~4.7 in~\cite{Kurtz}
is written for the more general setting
of continuous state space Markov chains.
We modify just the notation in~\cite{Kurtz}
in order to state the result in a way which is more suited to our particular setting.
Theorem~4.7 in~\cite{Kurtz} can be applied if the following 
set of conditions is satisfied.
There exist sequences of positive numbers 
$(\a_m)_{m\geq1}$ and $(\e_m)_{m\geq1}$
such that
\begin{enumerate}
\vspace*{-15 pt}
\item $\displaystyle\lim_{m\to\infty}\a_m=\infty\ $
and
$\ \displaystyle\lim_{m\to\infty}\e_m=0$.

\item $\displaystyle\sup_{m\geq1}\,\sup_{x\in E_m}\,\a_m\sum_{y\in E_m}|y-x|p^m(x,y)<\infty$.

\vspace*{-5 pt}
\item $\displaystyle\lim_{m\to\infty}\,\sup_{x\in E_m}\,\a_m
\sum_
{y\in E_m:|y-x|>\e_m}
|y-x|p^m(x,y)=0$.

\hspace*{-30 pt}Define, for $m\geq 1$,
$F^m(x)=\displaystyle\a_m\sum_{y\in E_m}(y-x)p^m(x,y)$.

\vspace*{-5 pt}
\item $\displaystyle\lim_{m\to\infty}\,\sup_{x\in E_m}\,|F^m(x)-F(x)|=0$.

\item There exists a constant $M$ such that
$$\forall x,y\in E\,,\qquad 
|F(x)-F(y)|\,\leq\, M|x-y|\,.$$
\end{enumerate}

\begin{theorem}[Kurtz]\label{Kurtz}
Suppose that conditions 1--5 are satisfied.
Suppose further that we have the convergence of the initial conditions
$$\lim_{m\to\infty}X^m_0=x^0\,.$$
Then, for every $\d,T>0$, we have
$$\lim_{m\to\infty}\,P\bigg(
\sup_{0\leq t\leq T}\big|
X^m_{\lfloor \a_m t\rfloor}-x(t)
\big|>\d
\bigg)\,=\,0\,.$$
\end{theorem}

\section{Proof of theorem~\ref{main}}\label{Proof}
Let $(Z_n)_{n\geq 0}$
be the Moran process
defined in section~\ref{Models}.
Our aim is to apply theorem~\ref{Kurtz}
to the sequence of Markov chains
$\big( (Z_n/m)_{n\geq0},m\geq 1 \big)$.
We only need to find the appropriate sequences
$(\a_m)_{m\geq 1}$ and $(\e_m)_{m\geq 1}$
and verify that conditions~1--5 are satisfied in our setting.
For $m\geq 1$, let 
$\a_m=\l m$ and $\e_m=2/m$.
The sequences $(\a_m)_{m\geq 1}$ and $(\e_m)_{m\geq 1}$
obviously verify condition~1.
As for condition~2,
we have, for $z\in\pmn$
$$\sum_{z'\in\pmn}\Big|
\frac{z}{m}-\frac{z'}{m}
\Big|p(z,z')\,=\,
\sum_{i,j=1}^N \Big|
-\frac{e_i}{m}+\frac{e_j}{m}
\Big|p(z,z-e_i+e_j)\,\leq\,\frac{\sqrt{2}}{m}\,.$$
Thus,
$$\sup_{m\geq 1}\,\sup_{z\in\pmn}\,\a_m\sum_{z'\in\pmn}
\Big|\frac{z'}{m}-\frac{z}{m}\Big|p(z,z')\,\leq\, \l\sqrt{2}\,<\,\infty\,,$$
as required for condition~2.
Since $p(z,z')>0$
if and only if 
$|z-z'|\leq \sqrt{2}$,
for all $m\geq 1$ and $z\in\pmn$, we have
$$\sum_{z'\in\pmn:|z'-z|>m\e_m}
\Big|
\frac{z'}{m}-\frac{z}{m}
\Big|p(z,z')\,=\,0\,,$$
and condition~3 is also satisfied.
Let $F:\R^N\to\R^N$
be the function defined by
$$\forall i\in\lbrace\,1,\dots,N\,\rbrace\,,
\quad\forall x\in\R^N\,,\qquad 
F_i(x)\,=\,\sum_{j=1}^N f_jQ_{ji}x_j-x_i\sum_{j=1}^N f_jx_j\,.$$
Since all the partial derivatives of $F$
are bounded on the simplex $\cS_N$,
F is a Lipschitz function on $\cS_N$,
i.e., condition~5 holds.
Finally,
let us compute,
for $m\geq 1$ and $x\in\pmn/m$,
the value of $F^m(x)$.
By definition,
$$F^m(x)\,=\,
\l m\sum_{z\in\pmn}\Big(
\frac{z}{m}-x
\Big)p(mx,z)
\,=\,\l\sum_{i,j=1}^N (-e_i+e_j)p(mx,mx-e_i+e_j)\,.
$$
Thus,
for $i\in\lbrace\,1,\dots,N\,\rbrace$
we have
\begin{align*}
F^m_i(x)\,&=\,
\l\sum_{k:k\neq i}p(mx,mx-e_k+e_i)
-\l\sum_{k:k\neq i}p(mx,mx-e_i+e_k)\\
&=\,\sum_{k:k\neq i}x_k\sum_{j=1}^N f_jQ_{ji}x_j
-\sum_{k:k\neq i}x_i\sum_{j=1}^N f_jQ_{jk}x_j\,.
\end{align*}
Since $x_1+\cdots+x_N=1$ and for all $i\in\lbrace\,1,\dots,N\,\rbrace$,
$Q_{i1}+\cdots+Q_{iN}=1$, 
$$F^m_i(x)=
(1-x_i)\sum_{j=1}^N f_jQ_{ji}x_j-x_i\sum_{j=1}^N f_j(1-Q_{ji})x_j
=\sum_{j=1}^N f_jQ_{ji}x_j-x_i\sum_{j=1}^N f_jx_j.
$$
Thus,
the function $F^m$ coincides with the function $F$
on the set $\pmn/m$, which readily implies condition~4.
Since all five conditions are satisfied,
we can apply theorem~\ref{Kurtz}
to the sequence of Markov chains $\big(
(Z_n/m)_{n\geq 0}, m\geq 1
\big)$
and we obtain the desired result.

\bibliographystyle{plain}
\bibliography{conveig}
\end{document}